\documentclass[a4paper,useAMS,usenatbib]{mn2e}

\usepackage{times}
\usepackage{graphicx}
\usepackage{psfig}

\begin{document}

 \title[Galactic rotation curves and brane world models]
 {Galactic rotation curves and brane world models}

 \author[F. Rahaman, M. Kalam, A. DeBenedictis, A. A. Usmani \& Saibal Ray]
 {F. Rahaman$^{1}$\thanks{E-mail: farook\_rahaman@yahoo.com},
M. Kalam$^{2}$\thanks{E-mail: mehedikalam@yahoo.co.in}, A.
DeBenedictis$^{3}$\thanks{E-mail: adebened@sfu.ca}, A. A.
Usmani$^{4}$\thanks{E-mail: anisul@iucaa.ernet.in},
Saibal Ray$^{5}$\thanks{E-mail: saibal@iucaa.ernet.in}\\
  $^{1}$Department of Mathematics, Jadavpur University, Kolkata 700 032,
 West Bengal, India\\
 $^{2}$Department of Physics, Netaji Nagar College for Women, Regent Estate, Kolkata
700 092, West Bengal, India\\ $^{3}$Pacific Institute for
Mathematical Sciences, Simon Fraser University Site \& \\
Department of Physics, Simon Fraser University, Burnaby, British
Columbia, V5A 1S6, Canada\\ $^{4}$Department of Physics, Aligarh
Muslim University, Aligarh 202 002, Uttar Pradesh, India \\
$^{5}$Department of Physics, Barasat Government College, North 24
Parganas, Kolkata 700 124, West Bengal, India }

\date{Accepted . Received ; in original form }

\pagerange{\pageref{firstpage}--\pageref{lastpage}} \pubyear{2008}

\maketitle

\begin{abstract}
In the present investigation flat rotational curves of the
galaxies are considered under the framework of brane-world models
where the $4d$ effective Einstein equation has extra terms which
arise from the embedding of the $3$-brane in the $5d$ bulk. It has
been shown here that these long range bulk gravitational degrees
of freedom can act as a mechanism to yield the observed galactic
rotation curves without the need for dark matter. The present
model has the advantage that the observed rotation curves result
solely from well-established non-local effects of gravitation,
such as dark radiation and dark pressure under a direct use of the
condition of flat rotation curves and does not invoke any exotic
matter field.
\end{abstract}


\section{Introduction}
The presence of dark matter was suspected for the first time by
\citet{Oort1930a,Oort1930b,Oort1930c} while carrying out his
observations of stellar motions in the local galactic
neighborhood. On the other hand, \citet{Zwicky1933,Zwicky1937}
discovered the presence of dark matter on a much larger scale
through his studies of galactic clusters. He, by comparing the
virial mass and luminous mass of galactic clusters, concluded that
a large amount of matter remains hidden in the galactic haloes.
Afterwards, observation of the flatness of galactic rotation
curves
\citep{Roberts1973,Ostriker1974,Einasto1974,Rubin1978,Rubin1979}
supported Zwicky's suggestion regarding the presence of
non-luminous matter. The flatness of the galactic rotation curves
are indicative of the presence of much more matter within the
galaxy than the visible material. Best estimates support a
spherical distribution of matter thus leading to the picture of a
visible galaxy surrounded by a dark matter halo. It can be
observed that dark matter gravitational effects are more manifest
at larger radius \citep{Begeman1989}.

It is seen that inside the optical radius the shapes of the
rotation curves probably correlate with the light distribution
\citep{Kent1987} whereas some degree of velocity variation has
been observed in the outer regions which might be related to the
overall mass distribution \citep{Albada1986}. The velocity
decrease may be even more than $50$ km/s on both sides of some of
the galaxies \citep{Casertano1991}. However, the measurements on
rotation curves in spiral galaxies show that the coplanar orbital
motion of gas in the outer parts of these galaxies keeps more or
less a constant value up to several luminous radii
\citep{Persic1996}. Another important feature is that the rotation
curve profile of a spiral galaxy is flat outside a central
galactic region \citep{Rubin1979} and this flatness of the
rotation curves might be due to the energy density of dark matter
which varies as $1/r^2$ \citep{Matos2000b}.

In subsequent years, gravitational lensing of objects like bullet
clusters and the temperature distribution of hot gas in galaxies
and galactic clusters as well as some other approaches have
further confirmed the existence of dark matter
\citep{Maoz1994,Cheng1999,Weinberg2002,Faber2006,Metcalf2007}.

Now, observation of anisotropy in the CMB \citep{Spergel2003}
suggests that the total energy-density of the Universe is $1.02
\pm 0.02$. On the other hand, observations of CMB and deuterium
abundance \citep{Sahni2004} as well as theoretical predictions
\citep{Olive2000} confirm that the baryon density of the Universe
cannot be more than $4 \%$ of the total energy-density. These
results hinted at a huge discrepancy between $\Omega_{total}$ and
$\Omega_{baryons}$. The introduction of Inflationary Theory by
\citet{Guth1981} and others \citep{Linde1982,Albrecht1982} enabled
theoreticians to conclude that the Universe must appear to be flat
and the total energy-density of the Universe is close to the
critical value. Since it was well established that visible matter
could not contribute more than $4-5 \%$ of the total
energy-density, it was initially speculated that $90-96 \%$ of the
total energy-density must be hidden matter, i.e. dark matter
\citep{Guzman2003}. But, in spite of intense searches, the
requisite amount of dark matter was not found. So, a crisis
developed in the cosmological arena. However, the emergence of the
idea of dark energy which is thought to be responsible for
accelerating the Universe \citep{Riess1998,Perlmutter1998} seems
to have, ultimately, removed that crisis. It is now known that
about two-thirds of the total energy-density comes in terms of
this dark energy while the remaining one-third is contributed by
matter, both visible and dark \citep{Sahni2004}. Thus, at present,
the dynamics of the Universe is governed by two components, the
exact nature of both of which are unknown, one is dark matter and
another is dark energy.

It has been argued that dark matter played a significant role
during structure formation in sub-megaparsec scales of the early
Universe. However, though the exact composition of dark matter is
still unknown, it is thought that most probably this is of
non-baryonic type. The reason behind this is that it is difficult
to reconcile baryonic dark matter with the small density
perturbations ($\delta\rho/\rho\sim 10^{-5}$ at $z\simeq 1100$)
measured by COBE and CMB experiments \citep{Sahni2004}. Moreover,
it is now quite established that the hot dark matter - like light
neutrino{\bf s} cannot contribute significantly to the large amount of
dark matter residing in the Universe, because in that case
galactic clusters with huge masses ($\sim 10^{15}M_{\odot}$) were
the first objects to form. However, this is not supported by the
observational constraint regarding neutrino mass and their density
\citep{Elgaroy2002,Minakata2003,Spergel2003,Ellis2003}. On the
other hand, cold dark matter, which does not contain any
appreciable internal thermal motion, can cluster on small scales
\citep{Sahni2004}. This supports the well known hierarchical
structure formation and suggests that most of the dark matter must
be cold and non-baryonic. The previous cold dark matter models,
introduced in the early $1980$'s, assumed that $\Omega_{CDM}=1$
and were known as Standard Cold Dark Matter (SCDM) models. In
spite of initial success SCDM models have fallen out of grace
\citep{Efstathiou1990,Pope2004}. Particularly, after the
introduction of the idea of an accelerating Universe, the SCDM
model is replaced by $\Lambda$-CDM (LCDM) model for including dark
energy as a part of the total energy density of the Universe. This
$\Lambda$-CDM model is found to be in nice agreement with various
sets of observations \citep{Tegmark2004a,Tegmark2004b}.

However, it is to be noted that not only the neutrino, as
mentioned earlier, but also many other candidates exist for dark
matter. Literature surveys reveal that \citet{Guzman2003}
considered quintessence-like dark matter in spiral galaxies
whereas Nukamendi and others
\citep{Nukamendi2000,Matos2000a,Matos2000b,Nukamendi2001,Lee2004}
have suggested that monopoles could be the galactic dark matter
whose energy density varies as $1/r^2$. However,
\citet{Nukamendi2000,Nukamendi2001,Lee2004,Rahaman2007} studied
global monopoles as a candidate for galactic dark matter in the
framework of the scalar tensor theory of gravity. In this line of
investigations \citet{Matos2000a,Matos2000b} have examined the
possibility and hence the type of dark matter that determines the
geometry of a spacetime where the flat rotational curves could be
explained. \citet{Cembranos2003} have shown that in the context of
brane-world scenarios with low tension massive brane fluctuations,
i.e. branons, are natural dark matter candidates and they could
make up the galactic halo. There are also examples where the
neutralino \citep{Nihei2005} and axino \citep{Panotopoulos2005}
dark matter have been studied in brane-world cosmology. Under the
same brane-world scenario it has been concluded that the neutral
hydrogen clouds at large distances from the galactic center may be
explained by postulating the existence of dark matter
\citep{Harko2005,Harko2006}. On the other hand, it is found by
\citet{Panotopoulos2007} that the gravitino can play the role of
dark matter in the Randall-Sundrum type II brane model
\citep{RS1999b} and determined what the gravitino mass should be
for different values of the five-dimensional Planck mass.
Interestingly, \citet{ref:yuri} have discussed the subject in
frames of the brane-world with induced curvature, which can
explain not only dark matter in galaxies but also dark matter
effects on the cosmological scale.

Therefore, it can be observed that in the framework of brane
cosmology researchers have made several attempts to understand
various features of galactic dark matter with different approaches
\citep{Mukohyama2000,Cembranos2003,Gumjudpai2003,Harko2005,Harko2006,Bohmer2007,Panotopoulos2007,ref:yuri}.

Here we will study a five dimensional brane-world model and show
that the long range bulk gravitational degrees of freedom can act
as a mechanism to yield the observed galactic rotation curves
without the need for dark matter. This model has the advantage
that the observed rotation curves result solely from
well-established non-local gravitational effects, such as dark
radiation and dark pressure, due to the $5d$ bulk and does not
rely on the introduction of an unknown and unseen matter field.

The above result that the dark radiation and its associated mass,
known as dark mass, being a linearly increasing function of the
distance has a similar behaviour as the dark matter at the
galactic scale have also been obtained by \citet{Bohmer2007}.
However, the approach of Harko and his collaborators is quite
different from us as their method is either involved in the
conformally symmetric space-times \citep{Mak2004,Harko2005} or
although they also consider constant rotational velocity with
different approach, their emphasis is on the qualitative analysis
of varied physical parameters in terms of observable quantities
\citep{Harko2006,Bohmer2007}. Unlike the previous studies the
motivation in the present study is primarily concerned with the
condition for flat rotation curves (which arises from the demand
that the considered test particles should have constant tangential
velocity) which has been exploited as a key to our entire
investigation. This simple supposition provides us the required
connection between dark matter and brane-world model in a very
straight forward way.

The scheme of the present investigation is as follows. In section
$2$ we present the relevant vacuum field equations for five
dimensional brane-world models whereas solutions will be sought
for and utilised to provide acceptable galactic rotation curves on
the brane in section $3$. We discuss in section $4$ how different
features of non-local effects of the brane-world scenarios are
relevant for dark matter and hence may be responsible for the
flatness of rotation curves which are in agreement with
observations.

\section{Field equations for five dimensional brane-world models}

In the simplest brane world models, a five dimensional space-time
is governed by the five dimensional Einstein field equations with
5d cosmological constant, $\Lambda_{5}$, which are given by
\begin{equation}
R^{\mu}_{\;\nu}-\frac{1}{2}R\,\delta^{\mu}_{\;\nu} -
(k_{5})^{2}\Lambda_{5} \delta^{\mu}_{\;\nu}= -(k_{5})^{2}
T^{\mu}_{\;\nu},
\end{equation}
where Greek indices take the values $(0,\,1,\,2,\,3,\,4)$,
$(k_{5})^{2}=8\pi G_{5}$ and $T^{\mu}_{\;\nu}$ is the
stress-energy tensor. Since matter is confined to the four
dimensional brane and only gravity permeates the bulk,
$T^{\mu}_{\;\nu}$ can therefore be written as
\begin{equation}
T^{\mu}_{\;\nu}\delta(Y)\left[T^{\mu}_{\;\nu\;(\mbox{\tiny{matter}})} +
\lambda_{b} \delta^{\mu}_{\;\nu}\right],
\end{equation}
where $Y=0$ denotes the location of the brane in the bulk and
${\lambda}_b$ is the vacuum energy on the brane (related to the
brane tension).

Since observations indicate that the dark matter distribution
around galaxies is spherical, we shall consider a static
spherically symmetric line element on the four-dimensional brane
of the form~\footnote{Here we are not concerned with the form of
the full five dimensional metric and therefore it is not
specified. Given a well behaved four dimensional metric, the bulk
metric's existence as a solution of the bulk field equations is
guaranteed by the Campbell-Maagard embedding theorem (Seahra \&
Wesson 2003; Wesson 2005).}
\begin{equation}
ds^2 = - e^{\nu(r)} dt^2 + e^{\lambda(r)} dr^2 + r^2 d \Omega ^2
\label{eq:4dmetric}
\end{equation}
with $d \Omega ^2=d \theta ^2 + sin^2 \theta d\phi^2$. Here $\nu$
and $\lambda$ are the metric potentials and are function of the
space coordinate $r$ only, such that $\nu=\nu(r)$ and
$\lambda=\lambda(r)$. We are making the reasonable assumption that
the rotation is not large enough to spoil the spherical symmetry
sufficiently to invalidate the metric (3).

The Gauss equations yield the following effective four-dimensional
gravitational field equations on the brane
\begin{eqnarray}
{G^{i}}_{j} ={R^{i}}_{j} - \frac{1}{2}{{g^{i}}_{j}} R
-\Lambda\,g^{i}_{\;j} = -8\pi {T^i}_j + (k_{5})^{2} S^{i}_{\;j} +
E^{i}_{\;j}, \label{eq:4dfieldeqs} \\
 \qquad (i,j = 0, 1, 2, 3)\,, \nonumber
\end{eqnarray}
with $S^{i}_{\;j}$ a quadratic function of the stress-energy
tensor (therefore $S^{i}_{\;j}=0$ in vacuum) and
$\Lambda=(k_{5})^{2}\left[\Lambda_{5} +
(k_{5})^{2}(\lambda_{b})^{2}/6 \right]/2$ is the $4d$ cosmological
constant which has been neglected in our calculations for
simplicity. Here $E^{i}_{\;j}$ is due to the long range
gravitational degrees of freedom and is a projection of the bulk
Weyl tensor onto the brane via $E_{\mu\nu}=C_{\mu \alpha \nu
\beta}n^{\alpha}n^{\beta}$ with $C$ the five dimensional Weyl
tensor and the $n$ vectors are five dimensional unit normal
vectors of the brane. Note from (\ref{eq:4dfieldeqs}) that on the
brane we experience standard $4d$ general relativity with the
exception that the source of the $4d$ Einstein tensor is augmented
by higher order stress-energy effects and $5d$ Weyl tensor
gravitational terms.

In vacuum, for the metric (\ref{eq:4dmetric}), equations
(\ref{eq:4dfieldeqs}) can explicitly be written as \citep{Mak2004}
\begin{equation}
e^{-\lambda} \left( \frac{\lambda^{\prime}}{r} - \frac{1}{r^{2}}
\right) + \frac{1}{r^{2}} = \frac{48\pi G}{k^4{\lambda}_b} U,
\end{equation}
\begin{equation}
e^{-\lambda} \left( \frac{\nu^{\prime}}{r} + \frac{1}{r^{2}}
\right) - \frac{1}{r^{2}} = \frac{16\pi G}{k^4{\lambda}_b} (U+2P),
\end{equation}
\begin{equation}
e^{-\lambda} \left (\nu^{{\prime}{\prime}} +
\frac{{\nu^{\prime}}^{2}}{2} - \frac{{\nu^{\prime}
\lambda^{\prime}}}{2} + \frac{\nu^{\prime} - \lambda^{\prime}}
{r}\right)= \frac{32\pi G}{k^4{\lambda}_b} (U-P),
\end{equation}
\begin{equation}
\nu^{\prime}=\frac{1}{2U+P}\left[U^{\prime}+2P^{\prime}+\frac{6P}{r}\right],
\end{equation}
where $U$ and $P$ are the dark radiation energy density (or simply
dark radiation) and dark pressure, respectively, of the bulk which
are function of the space coordinate $r$ only. According to
\citet{Mak2004} these parameters $U=U(r)$ and $P=P(r)$ are the
components of $f_{\mu\nu}$ which consists of the projection of the
bulk Weyl tensor on to the brane. This projected Weyl tensor
effectively serves as an additional matter source. Here a
$prime$ denotes derivative with respect to space coordinate $r$.

\section{Solutions for dark matter parameters on brane-world models}

Let us now consider the case of flat rotation curves for which the
required condition \citep{Rahaman2007} can be put as
\begin{equation}
e^{\nu}= B r^l, \label{eq:metricform}
\end{equation}
where $l=2{v_{\phi}}^{2}$ and $B$ is an integration constant.

\begin{figure}
\begin{center}
\vspace{0.5cm}
\includegraphics[bb=21 26 440 497, scale=0.5, clip, keepaspectratio=true]{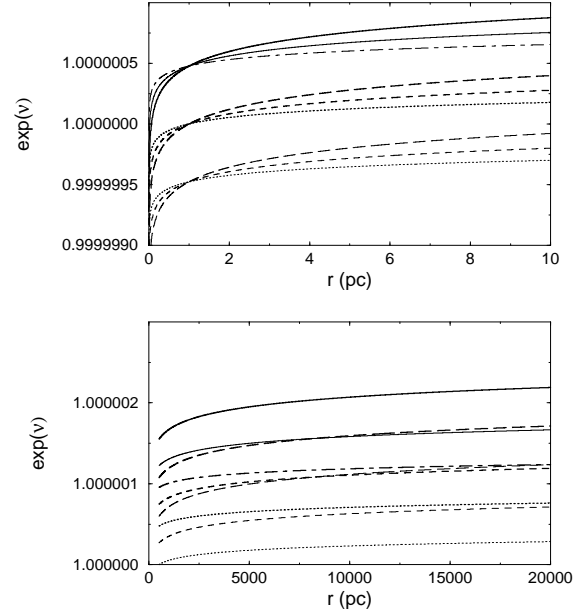}
\caption{Plot for the variation of $e^{\nu}$ vs $r$. The upper and
lower panels correspond to short and long $r$ behaviours. The
dotted, dashed and long-dashed curves represent $v_\phi=200$,
$250$ and $300$ Kms/second, respectively. For all these, thin
curves and thick curves represent $B=0.9999995$ and  $B=1.0$,
respectively. The chain, solid and thick solid curves
respectively, represent $v_\phi=200$, $250$ and $300$ Kms/second
but for $B=1.0000005$. } \label{fig:1}
\end{center}
\end{figure}

The form of (\ref{eq:metricform}) arises from demanding that test
particles, which obey the equation of motion \citep{Chandra1983}
\begin{equation}
v_{\phi}=v_{\mbox{\tiny{tangential}}}= \left[\frac{r
\left(e^{\nu}\right)^{\prime}}{2 e^{\nu}}\right]^{1/2} ,
\label{eq:tanvel}
\end{equation}
have constant tangential velocity. It is to be noted that the
observed rotational curve profile in the dark matter dominated
region is such that the rotational velocity $v_{\phi}$ becomes
more or less a constant with $v_{\phi} \sim 200 - 300$ km/s for a
typical galaxy
\citep{Binney1987,Persic1996,Matos2000a,Boriello2001}.

\begin{figure}
\begin{center}
\vspace{0.5cm}
\includegraphics[bb=10 86 502 654, scale=0.33, clip, height=5cm, width=5.3cm]{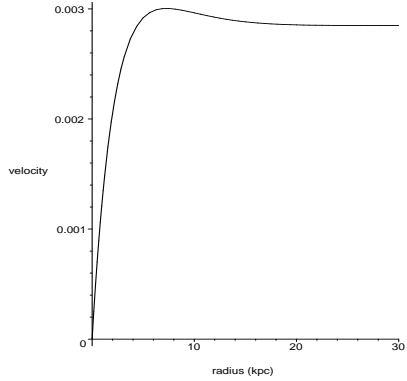}
\caption{A sample rotation curve utilising an ansatz equation for
$v_{\phi}$ of the form $v_{\phi}=\alpha
r\,\exp(-k_{1}r)+\beta\left[1-\exp(-k_{2}r)\right]$.}
\label{fig:2}
\end{center}
\end{figure}

One may generate an appropriate galactic rotation curve by fitting
to available observational data. An example of such a curve is
shown in figure \ref{fig:2}. The key point of interest is that the
curve must tend to an appropriate constant in the outer regions.
Equivalently, the metric function $e^{\nu(r)}$ must asymptotically
approach the form dictated by equation (\ref{eq:metricform}). A
reasonably general function which produces suitable rotation
curves for \emph{both} large and small $r$ is given by
\begin{equation}
 v_{\phi}=\alpha r\,\exp(-k_{1}r)+\beta\left[1-\exp(-k_{2}r)\right],
 \label{eq:theo_curve}
\end{equation}
where $\alpha$, $\beta$, $k_{1}$ and $k_{2}$ are constants determined by
appropriate fitting to data. Although our analysis will concentrate on
the outer regions, where the velocity is constant, a few comments
are appropriate for the inner regions.

\begin{figure}
\begin{center}
\vspace{0.5cm}
\includegraphics[bb=23 80 533 671, scale=0.4, clip, height=5cm, width=5.3cm]{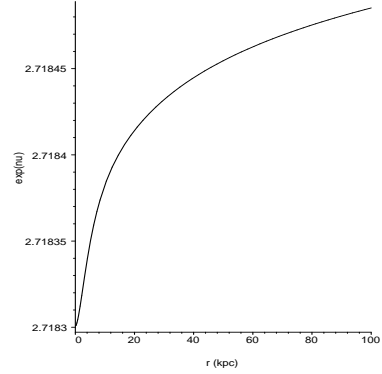}
\caption{The metric function $\exp[\nu(r)]$ generated by the
ansatz (\ref{eq:theo_curve}).} \label{fig:3}
\end{center}
\end{figure}

In the inner regions, it is expected that the mass due to the
concentration of stars will dominate the dynamics. A typical
galactic star will, however, be moving in what to a reasonable
approximation is vacuum, so that the vacuum equations are valid.
To make the analysis tractable we will assume a test-particle
star, located at some radius $r_{*}$, moving under the influence
of the spherically symmetric gravitational field due to stars
located inside its orbit ($r < r_{*}$) in the brane-world
scenario. Without detailed knowledge of the Weyl stresses one does
not know the exact form of the spherically symmetric vacuum
equations. However, given the form of the rotation curve as given
by the equation  (\ref{eq:theo_curve}), we can study the
properties the metric must possess. Equation (\ref{eq:tanvel}),
utilising (\ref{eq:theo_curve}), may actually be integrated to
yield an analytic expression for $\nu(r)$. The result is rather
unweildly and not very perspicuous so instead a plot of
$\mbox{e}^{\nu(r)}$ is provided in figure \ref{fig:3} from which
it may be verified that we have the asymptotic form of
(\ref{eq:metricform}) for appropriate $l$.

Now, from the equations (5) to (7) and then using the simple
proposition expressed in the equation (9) one obtains the
following simplified form
\begin{equation}
\frac{2}{r}=\left(2+l+\frac{l^2}{2}\right)\frac{e^{-\lambda}}{r}-
\left(2+\frac{l}{2}\right){\lambda}^{\prime} e^{-\lambda}.
\end{equation}
If we substitute $e^{-\lambda}=z$, then the above equation (12)
reduces to
\begin{equation}
z^{\prime}+\frac{az}{r}=\frac{2}{(2+\frac{l}{2})r},
\end{equation}
with $a=(2+l+\frac{l^2}{2})/(2+\frac{l}{2})$.

 This can be put in the following integral form
\begin{equation}
zr^a=\int\frac{2}{(2+\frac{l}{2})}r^{a-1} dr +D,
\end{equation}
where $D$ is an integration constant.

Therefore, after integration of the equation (14), we get the
metric potential $e^{-\lambda}$ as
\begin{equation}
e^{-\lambda}= \frac{2}{(2+\frac{l}{2})a}+\frac{D}{r^{a}}.
\end{equation}

\begin{figure}
\begin{center}
\vspace{0.5cm}
\includegraphics[bb=21 26 440 497, scale=0.5, clip, keepaspectratio=true]{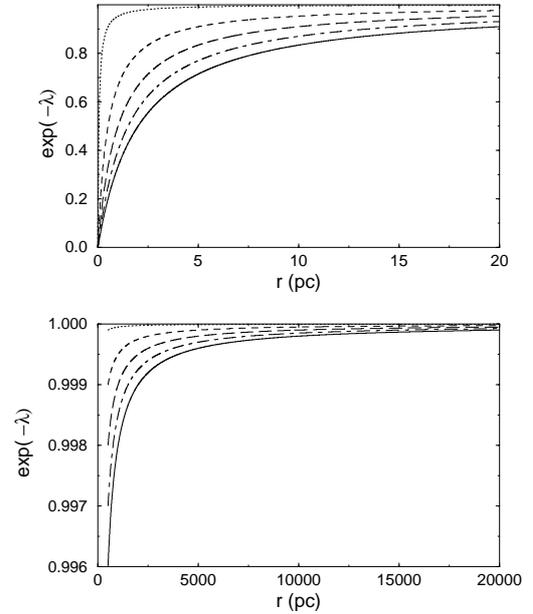}
\caption{Plot for the variation of $e^{-\lambda}$ vs $r$. The~
upper and lower panels correspond to short and long $r$
behaviours. The dotted, dashed, long-dashed, chain and solid
curves represent $D=0.05, 0.5, 1.0, 1.5$ and $2.0$, respectively.
} \label{fig:4}
\end{center}
\end{figure}

With the flat rotational curve condition, therefore, the metric
(3) becomes
\begin{equation}
ds^2 = - B_0 r^l dt^2 +
\left[\frac{2}{(2+\frac{l}{2})a}+\frac{D}{r^{a}}\right]^{-1} dr^2
+ r^2 d \Omega ^2.
\end{equation}

By the use of equation (15) in (5) one can easily obtain the
following expression for the dark radiation
\begin{equation}
U(r)=\frac{1}{b}\left[\frac{D(a-1)}{r^{a+2}}+\left(1-\frac{2}{a(2+\frac{l}{2})}\right)\frac{1}{r^2}\right]
\end{equation}
with $b=48\pi G/k^4{\lambda}_b$.

\begin{figure}
\begin{center}
\vspace{0.5cm}
\includegraphics[bb=14 26 440 497, scale=0.54, clip, keepaspectratio=true]{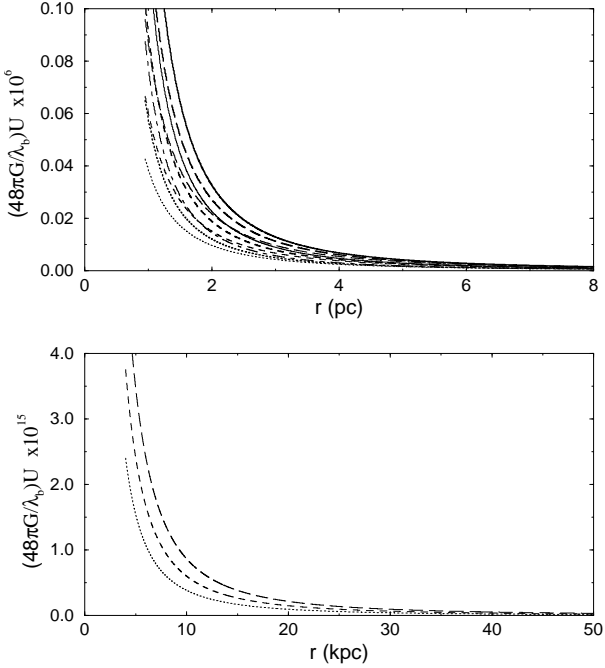}
\caption{Plot for the variation of $U$ vs $r$. The upper and lower
panels correspond to short and long $r$ behaviours. The dotted,
dashed and long-dashed curves represent $v_\phi=200$, $250$ and
$300$ Kms/second, respectively. For all these, thin curves and
thick curves correspond to $D=0.0$ and $D=1.0$, respectively. The
chain, solid and thick solid curves respectively, represent
$v_\phi=200$, $250$ and $300$ Kms/second but for $D=2.0$. In the
lower panel, plots represent the case for $D=1.0$ only because
results are not found significantly sensitive to $D$ at long
distances. } \label{fig:5}
\end{center}
\end{figure}

Now, after differentiating equation (9) and equating it with
equation (8) we get
\begin{equation}
\nu^{\prime}=\frac{l}{r}=-\frac{1}{2U+P}\left[U^{\prime}+2P^{\prime}+\frac{6P}{r}\right]
\end{equation}
which after substitution of equation (17) yields the following
first order differential equation
\begin{equation}
P^{\prime}+\frac{cP}{r}= \frac{d}{r^{a+3}}+\frac{e}{r^3}
\end{equation}
with  notations $c=(6+l)/2$,~$d=D(a-1)[a+2(1-l)]/2b$ and
$e=(1-l)\left(1-\frac{2}{a(2+\frac{l}{2})}\right)/b$.

\begin{figure}
\begin{center}
\vspace{0.5cm}
\includegraphics[bb=10 26 440 497, scale=0.54, clip, keepaspectratio=true]{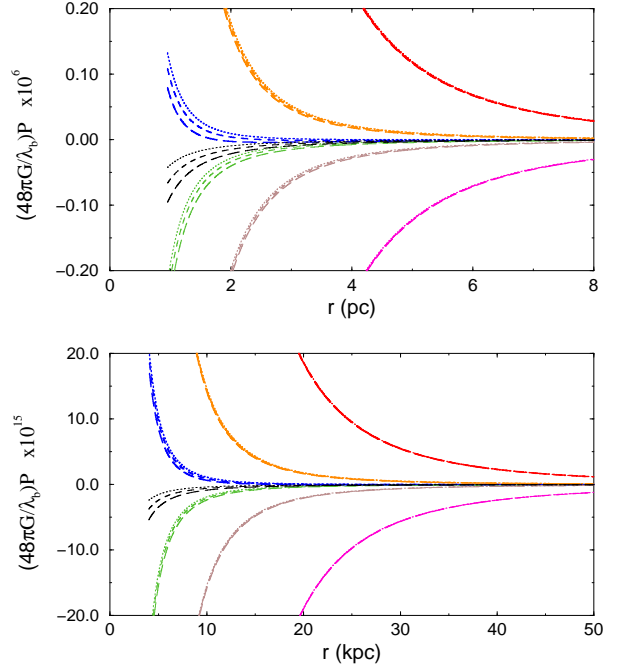}
\caption{Variation of $P$ vs $r$. The upper and lower panels
correspond to short and long $r$ behaviours. The black curves
represent $D=0.0$ for both the panel. For the upper panel, blue,
orange and red colours represent $D=0.0000001$, $D=0.000001$ and
$D=0.00001$, and green, brown and pink colours represent
$D=-0.0000001$, $D=-0.000001$ and $D=-0.00001$, respectively. For
the lower panel, blue, orange and red colours represent $D=0.001$,
$D=0.01$ and $D=0.1$, and green, brown and pink colours represent
$D=-0.001$, $D=-0.01$ and $D=-0.1$, respectively. For all the
colours, dotted, dashed and long-dashed curves correspond to
$v_\phi=200$, $250$ and $300$ Kms/second, respectively. }
\label{fig:6}
\end{center}
\end{figure}

 The above equation (19), after some manipulation, can be
written in the integral form as
\begin{equation}
Pr^{c}=\int \frac{d}{r^{3+a-c}} dr + \int \frac{e}{r^{3-c}} dr+E,
\end{equation}
where $E$ is an integration constant.

 The solution to the equation (20), related to dark pressure, can be given by
\begin{equation}
\label{eqpress}
P(r)=-\frac{d}{(2+a-c)r^{2+a}}-\frac{e}{(2-c)r^{2}}+\frac{E}{r^{c}}.
\end{equation}
 Here as $r$ tends to infinity $P$ will vanish. However, the
third term goes as $E/r^{3+l/2}$ so it seems to have the fall-off
properties more rapid than the other two terms. In this connection
it can also be observed that for large values of $l$, which is
proportional to square of the tangential velocity, both the $d$
term and the $E$ term possess similar fall-off properties. This
automatically suggests that for small values of tangential
velocity one can easily set $E$ to zero. Moreover, integration
constant $D\rightarrow 0$ implies $d\rightarrow 0$. With this, the
first term of equation (21) sets to zero too and only the second
term, $P(r)= -e/(2-c)r^{2}$, defines the pressure. We plot it in
Figure~$6$ as represented by black curves, wherein we observe that
pressure is always negative. The first term starts to dominate
even with a very small value of $D$ making the pressure positive
with positive value of $D$ as denominator of this term is
negative. The sensitivity of pressure with $D$ is plotted in
Figure~$6$. Curves for different rotational velocities are
distinguishable with $D=0$. However, they tend to overlap with
increasing value of $D$.

Let us now study an expression of active gravitational dark mass
associated with the dark radiation which can be given by
\begin{eqnarray}
M(r)=\int_{r_{min}}^r \frac{24\pi G}{k^4{\lambda}_b} U r^2 dr
\nonumber\\ =\frac{1}{2}\left[\left(1-\frac{2}{a(2+\frac{l}{2})}
\right)r - \frac{D}{r^{a-1}} \right]_{r_{min}}^r.
\end{eqnarray}
We observe from the equations (17) and (21) that $r=0$ gives a
singularity. Therefore, the lower limit of the integration does
not exist (at $r=0$). However this solution is valid only for some
radius $r > 0$, the interior regions being governed by the full
solution giving rise to (\ref{eq:theo_curve}), which yields well
behaved quantities at the origin. Here the equation (22) demands
the constraint to be imposed on $a$ is $a>1$ for any non-zero
value of $l$ as is evident from the expression of $a$ which is
$(2+l+\frac{l^2}{2})/(2+\frac{l}{2})$. The expression for dark
mass in the equation (22) clearly shows a radial dependence which
is increasing function of $r$ in the outer regions of this
analysis. We also observe that for an approximate value for the
integration constant, when $D \rightarrow 0$, this radial increase
gradually becomes more prominent. Our results have been shown in
the Figure~\ref{fig:6}. This effect is in accordance with the
experimental results in connection to the flatness of the galactic
rotation curves \citep{Begeman1989}.

\begin{figure}
\begin{center}
\vspace{0.5cm}
\includegraphics[bb=14 26 440 497, scale=0.54, clip, keepaspectratio=true]{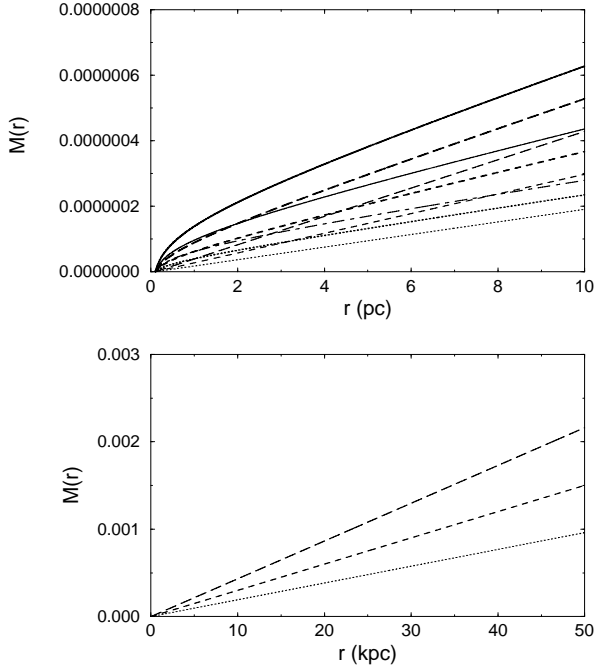}
\caption{Plot for the variation of $M(r)$ vs $r$. The upper and
lower panels correspond to short and long $r$ behaviours. The
dotted, dashed and long-dashed curves represent $v_\phi=200$,
$250$ and $300$ Kms/second, respectively. For all these, thin
curves and thick curves correspond to $D=0.0$ and $D=1.0$,
respectively. The chain, solid and thick solid curves
respectively, represent $v_\phi=200$, $250$ and $300$ Kms/second
but for $D=2.0$. In the lower panel, plots represent the case for
$D=1.0$ only because results are not found significantly sensitive
to $D$ at long distances. As equation for M is singular at $r=0$,
the minimum value of r is taken to be $0.1$ (pc). } \label{fig:7}
\end{center}
\end{figure}

\section{Discussions and Conclusions}
In this paper we have investigated the possibility of the
existence of dark matter via galactic rotation curves under the
framework of brane-world models. It has been observed that the
non-local effects of the bulk as manifested through dark radiation
and hence dark mass acts as the so-called dark matter which is
thought to be responsible for the flatness of rotation curves
around galaxies. From the equation (22) it can easily be seen that
$M$ is linearly increasing with the radial distance of the galaxy.
This is also clearly evident from the graphical plots (Figure~$1$
and Figures~$4$ - $7$) of the different physical quantities.
However, to get this connection we have employed a very simple
supposition as expressed in the equation~(\ref{eq:metricform})
which arises from the demand that test particles under
consideration will have constant tangential velocity. Note that
this constant tangential velocity actually refers to the flatness
of the galactic rotation curves.

In this context it can be observed that there is a striking
similarity between the Figures~$1$ and $4$ of the present work
under brane-world models with those of \citet{Rahaman2007} drawn
for a global monopole field in the Brans-Dicke theory under the
same equation~(\ref{eq:metricform}). Actually, \citet{Rahaman2007}
have shown by using Brans-Dicke theory that monopole could be the
galactic dark matter in the spiral galaxies whereas,
\citet{Guzman2000} have proposed scalar fields as dark matter in
the spiral galaxies~\footnote{However, in this context we would
like to make a point that global monopoles are a dark matter
candidate (with very strong restrictions) different to scalar
fields.}. This immediately indicates the global monopole as one of
the candidates for galactic dark matter irrespective of the
underlying theories. At the same time it seems that we get
confirmation of dark matter not only from the brane-world model
but also from Brans-Dicke theory. In this connection it is
asserted by \citet{Harko2005} that `` ... despite more than 20
years of intense experimental and observational effort, up to now
no {\it non-gravitational} evidence for dark matter has ever been
found ...''.

We would like to put some comments on the properties of Weyl
tensor which might be relevant here. It is already mentioned in
the field theoretical part that $E^{i}_{\;j}$ is due to the long
range gravitational degrees of freedom and is a projection of the
bulk Weyl tensor onto the brane via $E_{\mu\nu}=C_{\mu \alpha \nu
\beta}n^{\alpha}n^{\beta}$ with $C$ the five dimensional Weyl
tensor and the $n$ vectors are five dimensional unit normal
vectors of the brane. It is also mentioned in connection to
(\ref{eq:4dfieldeqs}) that on the brane, we experience standard
$4d$ general relativity with the exception that the source of the
$4d$ Einstein tensor is augmented by higher order stress-energy
effects and $5d$ Weyl tensor gravitational terms. It is therefore
highly probable, that if Randall-Sundrum models
\citep{RS1999a,RS1999b} are correct, the physical effects we
attribute to unseen material (for~example, dark matter and dark
energy) are actually due to these extra source terms. Another
point regarding the Weyl tensor is that the presence of Weyl
stresses means that the matching conditions do not have a unique
solution on the brane and hence knowledge of $5d$ Weyl tensor is
needed as a minimum condition for uniqueness. It is shown by
\citet{Dadhich2000} that in the $5d$ brane, the high energy
corrections to the energy density, together with the Weyl stresses
from bulk gravitons, imply that on the brane the exterior metric
of a static source is no longer the Schwarzschild metric.

There have been several studies addressing the issue of acceptable
spherically symmetric vacuum solutions in brane-world scenarios
\citep{ref:viswilt,ref:harkmakvac,ref:creek,ref:pdl,ref:yuri}. It
was explicitly demonstrated in these references how the
brane-world equations provide a weaker requirement on the $4d$
brane metric than the Einstein equations in four dimensions.
Therefore, a more general class of solutions may be admitted in
the brane-world picture than the Schwarzschild solution. Ponce de
Leon demonstrated that the non-static situation differs from the
static one by the extra requirement that $T^{0}_{\;0}=T^{1}_{\;1}$
at the matter-vacuum boundary. This requirement is sufficient to
close the system of equations, yielding a Schwarzschild or
Reissner-Nordstr\"{o}m-like metric (where the ``charge'' term
arises from an integration constant) as possible exteriors for
non-static systems. For the static system he showed that no such
restriction occurs at the boundary and therefore the vacuum
solutions admit a much wider (infinitely many) possible vacua.

\section*{Acknowledgments} Authors (AAU and SR) are
thankful to the authority of Inter-University Centre for Astronomy
and Astrophysics, Pune, India for providing them Associateship
programme under which a part of this work was carried out. We also
thank referee for his valuable suggestions which have enabled us
to improve the manuscript substantially.

{}

 \end{document}